\documentclass[aps,prl,reprint,twocolumn,preprintnumbers,superscriptaddress,letterpaper]{revtex4-2}
\usepackage[english]{babel}
\usepackage{ucs}
\usepackage[utf8x]{inputenc}
\usepackage{amsmath}
\usepackage{amsfonts}
\usepackage{amssymb}
\usepackage{graphicx}
\usepackage{bm}
\usepackage{bbm}
\usepackage{empheq}
\usepackage{xr-hyper}
\usepackage{hyperref}
\usepackage{xcolor}

\begin{document}
\title{Information Constraints for Scalable Control in a Quantum Computer}
\author{John M. Martinis}
\affiliation{Quantala, Santa Barbara, CA 93105, USA}
\email{martinis@quantala.tech}
\date{\today}

\begin{abstract}
When working to understand quantum systems engineering, there are many constraints to building a scalable quantum computer.  Here I discuss a constraint on the qubit control system from an information point of view, showing that the large amount of information needed for the control system will put significant constraints on the control system.  The size the qubits is conjectured to be an important systems parameter.    
\end{abstract}
\maketitle

Quantum computing has entered a compelling scientific era as quantum algorithms can now be run on multiple physical systems.  Crucially, builders can obtain feedback on how best to assemble a fully functional quantum computer and thus are developing strategies for quantum systems engineering.

Although quantum computation power has exceeded classical supercomputers for a quantum supremacy experiment\cite{qs}, a crafted problem, scientists still search for computational advantage in a generically useful algorithm\cite{NISQ}.  Such advances are limited by the noise and decoherence of present-day systems.
Thus, a major area of research is developing error-corrected quantum computers, where noise is measured and corrected with redundant encoding of qubit information, similar to classical error correction.  The surface code is presently a leading approach to quantum error correction\cite{scorigin,sc}, but a serious challenge is the large overhead of order 1000 physical qubits per logical qubit \cite{scaleup}.  Understanding a practical architecture and pathway for scaling to a large quantum computer is presently a fundamental question.

Although one can imagine how large chips can be made using integrated-circuit fabrication technology, a serious issue is how to control a large number of qubits; an estimate is 100\,k - 1\,M qubits for a classically intractable problem \cite{useful_plane,useful_molec}, which assumes $\sim 0.1\%$ for gate errors \cite{scaleup}. Here, I discuss the information complexity that is needed for the qubit control system. 

The main idea is that to reach small qubit errors for present systems, the controls need many parameters.  A quantum computer is calibrated by tuning these parameters.  For example in the quantum supremacy experiment, more than 100 parameters are set per qubit.  This large number of control parameters thus implies much information is needed to properly operate each qubit with low errors.  Such information can be stored either as the parameters or as the waveforms to be sent to the qubits.  Whatever the design of the control system, there must be a reasonable degree of complexity, volume, and expense in producing such finely tuned control.  This is in stark contrast to typical classical systems where the transistors are simply connected to the power supply.

In order to connect the control system to the qubits, it seems reasonable that the connections and controls should be somewhat smaller than the qubit size.  Of course, this comparison depends greatly on the particular technology and wiring details, but the basic conclusion is that large qubits are advantageous, since complex control systems are not small.  This again is in stark contrast to classical computing, where the small size of transistors is a fundamental advantage.  

For superconducting systems, the size of transmon qubits \cite{transmon} are about 1 mm.  This should be large enough to match wiring and control systems based on existing technology.  

It is important to mention that a large number of parameters is not fundamental to control in quantum computing.  In principle, a qubit only needs three parameters for a single qubit gate, the qubit frequency and the amplitude and phase of the control pulse.  This could be reduced to two parameters for systems with fixed transition frequencies like atoms.  This characteristic may enable multiplexed control systems, where control signals are sent to a large patch of qubits, say one logical qubit, and simpler individual qubit controls are used to turn the controls on and off, and fine tune the amplitude and phase of each qubit.  However, since present qubits are limited by gate errors, there is no engineering margin in qubit errors to trade-off against simplified control, so it is not clear this strategy is viable.  In time, however, such multiplexed control might become feasible. 

In summary, there are significant engineering constraints based on the large information content of qubit controls.  Complex control favors large qubits.


\begin{thebibliography}{10}
\bibitem{qs}
F. Arute \textit{et. al.}, 
Quantum supremacy using a programmable superconducting processor, 
Nature \textbf{574}, 505 (2019).
\bibitem{NISQ}
J. Preskill, 
Quantum computing in the NISQ era and beyond,
Quantum \textbf{2}, 79 (2018).
\bibitem{scorigin}
S. B. Bravyi and A. Y. Kitaev, arXiv:quant-ph/9811052.
\bibitem{sc}
A. G. Fowler, M. Mariantoni, J. M. Martinis, and A. N. Cleland, 
Surface codes: Towards practical large-scale quantum computation, 
Physical Review A \textbf{86}, 032324 (2012).
\bibitem{scaleup}
J. M. Martinis,
Qubit metrology for building a fault-tolerant quantum computer,
NPJ Quantum Information \textbf{1}, 15005 (2015).
\bibitem{useful_plane}
I. D. Kivlichan, C. Gidney, D. W. Berry, N. Wiebe, J. McClean, W. Sun, Z. Jiang, N. Rubin, A. Fowler, A. Aspuru-Guzik, H. Neven, and R. Babbush,
Improved Fault-Tolerant Quantum Simulation of Condensed-Phase Correlated Electrons via Trotterization, 
Quantum \textbf{4}, 296 (2020).
\bibitem{useful_molec}
D. W. Berry, C. Gidney, M. Motta, J. R. McClean, and R. Babbush,
Qubitization of Arbitrary Basis Quantum Chemistry Leveraging Sparsity and Low Rank Factorization, Quantum \textbf{3}, 208 (2019).
\bibitem{transmon}
J. Koch \textit{et al.},
Charge-insensitive qubit design derived from the Cooper pair box,
Phys. Rev. A \textbf{76}, 042319 (2007).

\end{thebibliography}
\end{document}